\documentclass[12pt]{article}
\usepackage{epsfig}
\begin{document}
\title{Photoproduction of $\Lambda(1520)$ hyperons from nuclei\\
 near the threshold }
\author{E. Ya. Paryev\\
{\it Institute for Nuclear Research, Russian Academy of Sciences,}\\
{\it Moscow 117312, Russia}}

\renewcommand{\today}{}
\maketitle

\begin{abstract}
   We present a theoretical investigation of $\Lambda(1520)$ attenuation in ${\gamma}A$ reactions
   near the threshold. It is performed in the framework of a collision model based on
   the nuclear spectral function. The model accounts for both primary photon--nucleon
   ${\gamma}N \to K\Lambda(1520)$ and secondary pion--nucleon ${\pi}N \to K\Lambda(1520)$
   production processes. We calculate the target mass and momentum dependences of the forward
   $\Lambda(1520)$ hyperon production from nuclei at photon energy of 2 GeV as well as for two
   options for its in-medium width. We find that the considered dependences are markedly
   sensitive to this width. Our studies also demonstrate that the secondary channel
   ${\pi}N \to K\Lambda(1520)$ plays a substantial role in the intermediate momentum
   $\Lambda(1520)$ photoproduction on nuclei in the chosen kinematics
   and, hence, is to be taken into account in the analysis of
   $\Lambda(1520)$ hyperon photoproduction from nuclei with
   the aim to get information on its width in the matter.
\end{abstract}

\newpage

\section*{1. Introduction}

\hspace{1.5cm} The study of the modifications of the hadronic spectral functions in nuclear
matter is one of the most attractive topics in modern nuclear physics. It has been motivated
by the expectation to get valuable information about the change (reduction) of scalar two-quark
condensate in the strongly interacting environment. The observation of such change could be
interpreted as evidence for the partial restoration of chiral symmetry at finite density.
The production of mostly light vector mesons $\rho$, $\omega$, $\phi$ in nuclear reactions with
photons [1--9] and protons [10--12] as well as in heavy--ion collisions [13--19] has been studied
to look for possible renormalizations of their spectral functions in nuclear medium through the
invariant-mass and transparency ratio measurements. The latter ones provide an access to the
in-medium meson width, which determines the attenuation of meson flux in nuclei and with it the meson
production yield from them. In the low-density approximation, this width can be related to the
in-medium meson--nucleon total cross section. A recent overview of theoretical and experimental
activities in this field is given in [20]. An almost all experiments reported a substantial
in-medium broadening of the $\rho$, $\omega$, $\phi$ spectral functions, whereas
the shift of their pole masses in nuclear matter was found only in the KEK experiment [10, 11].
A very recent experimental results
on the transparency ratio for the photoproduced $\eta^{\prime}$ mesons are presented in [21].
They show that the in-medium width of the $\eta^{\prime}$ is approximately equal to 25 MeV.

  The another interesting case of resonance medium modification
is that of the $\Lambda(1520)$ hyperon where the hadronic model [22] predicts
an increase of the width of low-momentum $\Lambda(1520)$ hyperons at saturation density by a factor of five
compared to its vacuum value ($\Gamma_{\Lambda(1520)}=15.6$ MeV) due to the medium modifications of their ${\bar K}N$
and ${\pi}\Sigma$ decay channels as well as the existence of the $\Lambda(1520) \to {\pi}\Sigma(1385)$ decay mode.
This mode is closed in the free space for the nominal masses of the $\Lambda(1520)$ and $\Sigma(1385)$,
but opens in the nuclear medium when the $\pi$ is allowed to become a $p$--$h$ excitation.
Whereas the mass shift of the
$\Lambda(1520)$ hyperon in nuclear matter is found to be small (about 2\% of its free mass at normal nuclear matter density $\rho_0$). Apart from $\phi$ meson [20], the similar medium effects were predicted very recently
for the $\bar {K^*}$ meson [23]. The sizable change in the $\Lambda(1520)$ (or $\Lambda^*$) width in the medium
found in [22] could be observed experimentally and in [24] the nuclear transparency ratio for the
$\Lambda(1520)$ has been calculated in photon- and proton-induced reactions. The calculations account for only
the relevant elementary one-step $\Lambda(1520)$ production mechanisms and neglect
the $\Lambda(1520)$ creation in the two-step processes with an intermediate pions.
As was found in [25], these processes contribute distinctly
to the "low-momentum" $\Lambda(1520)$ creation in near-threshold $pA$ reactions and, hence, they
may contribute to the $\Lambda(1520)$ production in near-threshold ${\gamma}A$ interactions
and thus distort the transparency ratio of $\Lambda(1520)$ in these interactions.
To gain a better understanding of the possible impact of above two-step processes on the
$\Lambda(1520)$ hyperon yield in ${\gamma}A$ collisions, the further theoretical studies are needed.

The main aim of the present work is to extend the spectral function approach [25] that has been employed
by us for the study of inclusive $\Lambda(1520)$ production in proton--nucleus reactions to
$\Lambda(1520)$-producing electromagnetic processes. In this paper,
we investigate on the basis of model [25]
the A and momentum dependences of the absolute and relative $\Lambda(1520)$
production cross sections from ${\gamma}A$ reactions at incident photon energy of 2 GeV by considering
respective primary photon--nucleon and secondary pion--nucleon  $\Lambda(1520)$ production processes.

\section*{2. Primary $\Lambda(1520)$ production processes}

\hspace{1.5cm} A $\Lambda(1520)$ resonance can be produced directly in the first inelastic
${\gamma}N$ collision due to the nucleon's Fermi motion.
At near-threshold beam energy of 2 GeV of our interest,
the following elementary processes which have the lowest free production threshold ($\approx$ 1.69 GeV)
mainly contribute:
\begin{equation}
\gamma+p \to {K^+}+\Lambda(1520),
\end{equation}
\begin{equation}
\gamma+n \to {K^0}+\Lambda(1520).
\end{equation}
Using
the results given in [25], we can represent the inclusive cross section for the production on nuclei
$\Lambda(1520)$ hyperons with the momentum ${\bf p}_{\Lambda^*}$
from the primary photon--induced reaction channels (1), (2) as follows:
\begin{equation}
\frac{d\sigma_{{\gamma}A\to {\Lambda(1520)}X}^{({\rm prim})}
({\bf p}_{\gamma})}
{d{\bf p}_{\Lambda^*}}=I_{V}[A]
\end{equation}
$$
\times
\left[\frac{Z}{A}\left<\frac{d\sigma_{{\gamma}p\to K^+{\Lambda(1520)}}({\bf p}_{\gamma},M_{\Lambda^*},
{\bf p}_{\Lambda^*})}{d{\bf p}_{\Lambda^*}}\right>+
\frac{N}{A}\left<\frac{d\sigma_{{\gamma}n\to K^0{\Lambda(1520)}}({\bf p}_{\gamma},M_{\Lambda^*},{\bf p}_{\Lambda^*})}{d{\bf p}_{\Lambda^*}}\right>\right],
$$
where
\begin{equation}
I_{V}[A]=2{\pi}A\int\limits_{0}^{R}r_{\bot}dr_{\bot}
\int\limits_{-\sqrt{R^2-r_{\bot}^2}}^{\sqrt{R^2-r_{\bot}^2}}dz
\rho(\sqrt{r_{\bot}^2+z^2})
\end{equation}
$$
\times
\exp{\left[-\sigma_{{\gamma}N}^{{\rm tot}}A\int\limits_{-\sqrt{R^2-r_{\bot}^2}}^{z}
\rho(\sqrt{r_{\bot}^2+x^2})dx
-\int\limits_{z}^{\sqrt{R^2-r_{\bot}^2}}\frac{dx}
{\lambda_{\Lambda^*}(\sqrt{r_{\bot}^2+x^2},M_{\Lambda^*})}\right]},
$$
\begin{equation}
\lambda_{\Lambda^*}({\bf r},M_{\Lambda^*})=\frac{p_{\Lambda^*}}{M_{\Lambda^*}
\Gamma_{{\rm tot}}({\bf r},M_{\Lambda^*})}
\end{equation}
and
\begin{equation}
\left<\frac{d\sigma_{{\gamma}N\to K{\Lambda(1520)}}({\bf p}_{\gamma},M_{\Lambda^*},
{\bf p}_{\Lambda^*})}
{d{\bf p}_{\Lambda^*}}\right>=
\int\int
P({\bf p}_t,E)d{\bf p}_tdE
\left[\frac{d\sigma_{{\gamma}N\to K{\Lambda(1520)}}(\sqrt{s},M_{\Lambda^*},{\bf p}_{\Lambda^*})}
{d{\bf p}_{\Lambda^*}}\right],
\end{equation}
\begin{equation}
  s=(E_{\gamma}+E_t)^2-({\bf p}_{\gamma}+{\bf p}_t)^2,
\end{equation}
\begin{equation}
   E_t=M_A-\sqrt{(-{\bf p}_t)^2+(M_{A}-m_{N}+E)^{2}}.
\end{equation}
Here,
$d\sigma_{{\gamma}N\to K{\Lambda(1520)}}(\sqrt{s},M_{\Lambda^*},{\bf p}_{\Lambda^*}) /d{\bf p}_{\Lambda^*}$
are the off-shell
differential cross sections for $\Lambda(1520)$ production in reactions (1) and (2)
at the ${\gamma}N$ center-of-mass energy $\sqrt{s}$;
$\sigma_{{\gamma}N}^{{\rm tot}}$ is the total cross section
of free ${\gamma}N$ interaction
($\sigma_{{\gamma}N}^{{\rm tot}}=0.15$ mb for photon energy of 2 GeV [26]);
${\bf p}_{\gamma}$ and $E_{\gamma}$ are the momentum and total energy of the initial photon,
and the definition of other quantities, entering into (3)--(8), is given in [25].

Following ref. [25], we assume that the off-shell differential cross sections \\
$d\sigma_{{\gamma}N\to K{\Lambda(1520)}}(\sqrt{s},M_{\Lambda^*},{\bf p}_{\Lambda^*}) /d{\bf p}_{\Lambda^*}$
for $\Lambda(1520)$ production in the reactions (1), (2), entering into eqs. (3), (6),
are equivalent to the respective on-shell cross sections calculated for the off-shell kinematics of the elementary processes  (1), (2). Taking into consideration the two-body kinematics of these processes, we can
readily get the following expressions for the former ones:
\begin{equation}
\frac{d\sigma_{{\gamma}N\rightarrow K\Lambda(1520)}(\sqrt{s},M_{\Lambda^*},{\bf p}_{\Lambda^*})}
{d{\bf p}_{\Lambda^*}}
={\frac{{\pi}}{I_2(s,m_K,M_{\Lambda^*})E_{\Lambda^*}}}
{\frac{d\sigma_{{\gamma}N\rightarrow K\Lambda(1520)}({\sqrt{s}},{\theta_{\Lambda^*}^*})}
{d{\bf \Omega}_{\Lambda^*}^{*}}}\times
\end{equation}
$$
\times
{\frac{1}{(\omega+E_t)}}\delta\left[\omega+E_t-\sqrt{m_K^2+({\bf Q}+{\bf p}_t)^2}\right],
$$
where
\begin{equation}
I_2(s,m_K,M_{\Lambda^*})=\frac{\pi}{2}\frac{\lambda(s,m_{K}^{2},M_{\Lambda^*}^{2})}{s},
\end{equation}
\begin{equation}
\lambda(x,y,z)=\sqrt{{\left[x-({\sqrt{y}}+{\sqrt{z}})^2\right]}{\left[x-
({\sqrt{y}}-{\sqrt{z}})^2\right]}},
\end{equation}
\begin{equation}
\omega=E_{\gamma}-E_{\Lambda^*}, \,\,\,\,{\bf Q}={\bf p}_{\gamma}-{\bf p}_{\Lambda^*}.
\end{equation}
Here,
$d\sigma_{{\gamma}N\rightarrow K\Lambda(1520)}({\sqrt{s}},{\theta_{\Lambda^*}^*})/d{\bf \Omega}_{\Lambda^*}^{*}$ are the off-shell differential cross sections for the production of a
$\Lambda(1520)$ hyperon under the polar angle ${\theta_{\Lambda^*}^*}$ in the ${\gamma}N$ c.m.s.;
$E_{\Lambda^*}$ is its total energy
($E_{\Lambda^*}=\sqrt{p_{\Lambda^*}^2+M_{\Lambda^*}^2}$) in l.s. and $m_K$ is the bare kaon mass.

    The recent detailed experimental information from SAPHIR Collaboration concerning the differential
cross section of the ${\gamma}p \to K^+\Lambda(1520)$ reaction
in the photon energy range $1.69~{\rm GeV}<E_{\gamma}<2.65$ GeV can be fitted as [27]
\footnote{It should be noted that at photon energies below 2.4 GeV the reaction ${\gamma}p \to K^+\Lambda(1520)$
has been recently studied also by the LEPS Collaboration at SPring-8 facility [28]. The data on
differential cross section of this reaction obtained in [28] are consistent, as the comparison shows,
with the SAPHIR results [27]. As far as the another experimental data on $\Lambda(1520)$ photoproduction
on the proton is concerned, there are only two old published results at higher energies of
$E_{\gamma}=$2.8--4.8 GeV [29] and $E_{\gamma}=11$ GeV [30] as well as new preliminary data in the
photon energy range of $E_{\gamma}=$1.75--5.5 GeV from the CLAS Collaboration [31]. The reaction
${\vec {\gamma}}p \to K^+\Lambda(1520)$ with linearly polarized photon beams has been very recently studied
again by the LEPS Collaboration at energies from the threshold to 2.4 GeV [32].}
:
\begin{equation}
\frac{d\sigma_{{\gamma}p\rightarrow K^+\Lambda(1520)}({\sqrt{s}},{\theta_{K}^*})}{dt}=
\left\{
\begin{array}{llll}
	2.53e^{-5.41|t-t^+|}
	&\mbox{for $1.69~{\rm GeV}< E_{\gamma} \le 1.93~{\rm GeV}$}, \\
	&\\
                   1.40e^{-2.20|t-t^+|}
	&\mbox{for $1.93~{\rm GeV}<E_{\gamma} \le 2.17~{\rm GeV}$}, \\
                   &\\
                   0.66e^{-1.63|t-t^+|}
                   &\mbox{for $2.17~{\rm GeV}<E_{\gamma} \le 2.43~{\rm GeV}$}, \\
                   &\\
                   0.49e^{-1.51|t-t^+|}
                   &\mbox{for $2.43~{\rm GeV}<E_{\gamma} \le 2.65~{\rm GeV}$}.
\end{array}
\right.	
\end{equation}
Here, $t$ is the squared four-momentum transfer between the incident photon and the outgoing $K^+$ meson,
$t^+$ is its maximal value which corresponds to the $t$ where the $K^+$ is produced at angle
$\theta_{K}^*=0^{\circ}$ in ${\gamma}p$ c.m.s. The differential cross section
$d\sigma_{{\gamma}p\rightarrow K^+\Lambda(1520)}/dt$ is measured in ${\rm {\mu}}$b/GeV$^2$.
In order to evaluate the invariant $t$ it is more convenient to put ourselves in the ${\gamma}p$ c.m.s.
Then, it can be expressed through the energies and momenta of the $K^+$ meson and the initial photon,
$E_{K}^*$, ${\bf p}_{K}^*$ and $E_{\gamma}^*$, ${\bf p}_{\gamma}^*$, in this system in the following way:
\begin{equation}
t=m_{K}^2-2E_{\gamma}^*E_{K}^*+2p_{\gamma}^*p_{K}^*\cos{\theta_{K}^*},
\end{equation}
where
\begin{equation}
p_{K}^*=|{\bf p}_{K}^*|=\frac{1}{2\sqrt{s}}\lambda(s,m_{K}^2,M_{\Lambda^*}^{2}),\,\,\,\,
E_{K}^*=\sqrt{{p_{K}^*}^2+m_{K}^2}
\end{equation}
and
\begin{equation}
p_{\gamma}^*=|{\bf p}_{\gamma}^*|=\frac{1}{2\sqrt{s}}\lambda(s,0,E_{t}^{2}-p_{t}^2),\,\,\,\,
E_{\gamma}^*=p_{\gamma}^*.
\end{equation}
If the struck target proton is on-shell, then in (16) $E_{t}^{2}-p_{t}^2=m_p^2$ ($m_p$ is the bare proton mass).
Otherwise, this quantity should be calculated in line with eq. (8).

   Using the relation (14), one finds that the quantity $|t-t^+|$, entering into eq. (13), reduces to a form:
\begin{equation}
|t-t^+|=2p_{\gamma}^*p_{K}^*(1-\cos{\theta_{K}^*}).
\end{equation}
The $\Lambda(1520)$ hyperon production angle ${\theta_{\Lambda^*}^*}$ in the ${\gamma}p$ c.m.s. is defined by:
\begin{equation}
\cos{\theta_{\Lambda^*}^*}=\frac{{\bf p}_{\gamma}^*{\bf p}_{\Lambda^*}^*}{p_{\gamma}^*p_{\Lambda^*}^*}
\end{equation}
and
\begin{equation}
\theta_{K}^*+\theta_{\Lambda^*}^*=\pi.
\end{equation}
Writting the invariant $u$--the squared four-momentum transfer between the incident photon and the secondary
 $\Lambda(1520)$ hyperon--in the l.s. and in the ${\gamma}p$ c.m.s. and equating the results, we readily obtain:
\begin{equation}
\cos{\theta_{\Lambda^*}^*}=\frac{p_{\gamma}p_{\Lambda^*}\cos{\theta_{\Lambda^*}}+
(E_{\gamma}^*E_{\Lambda^*}^*-E_{\gamma}E_{\Lambda^*})}
{p_{\gamma}^*p_{\Lambda^*}^*}.
\end{equation}
In the above, $\theta_{\Lambda^*}$ is the angle between the momenta ${\bf p}_{\gamma}$ and ${\bf p}_{\Lambda^*}$
in the l.s. frame. Differentiation the $t$ with respect to $\cos{\theta_{K}^*}$ permits us to get the
following relation between the differential cross section
$d\sigma_{{\gamma}p\rightarrow K^+\Lambda(1520)}({\sqrt{s}},{\theta_{\Lambda^*}^*})/d{\bf \Omega}_{\Lambda^*}^{*}$
for $\Lambda(1520)$ production in reaction (1), entering into eq. (9), and the cross section
$d\sigma_{{\gamma}p\rightarrow K^+\Lambda(1520)}({\sqrt{s}},{\theta_{K}^*})/dt$, parametrized by eq. (13):
\begin{equation}
\frac{d\sigma_{{\gamma}p\to K^+{\Lambda(1520)}}(\sqrt{s},\theta_{\Lambda^*}^*)}{d{\bf \Omega}_{\Lambda^*}^*}=
\frac{p_{\gamma}^*p_{\Lambda^*}^*}{\pi}
\frac{d\sigma_{{\gamma}p\rightarrow K^+\Lambda(1520)}({\sqrt{s}},{\theta_{K}^*})}{dt},
\end{equation}
where $\theta_{K}^*=\pi-\theta_{\Lambda^*}^*$. When the reaction ${\gamma}p\to K^+{\Lambda(1520)}$ goes on an
off-shell target proton, then instead of the incident photon energy $E_{\gamma}$, appearing in the eq. (13), we
should use the effective energy $E_{\gamma}^{\rm eff}$. It can be expressed in terms of collision energy
squared $s$, defined by eq. (7), as follows:
\begin{equation}
E_{\gamma}^{\rm eff}=\frac{s-m_p^2}{2m_p}.
\end{equation}

  For obtaining the differential cross section
$d\sigma_{{\gamma}n\rightarrow K^0\Lambda(1520)}({\sqrt{s}},{\theta_{\Lambda^*}^*})/d{\bf \Omega}_{\Lambda^*}^{*}$
of the reaction (2) at the near-threshold energies,
we have assumed in line with the preliminary experimental results of ref. [31] that there is the isospin symmetry
in the $\Lambda(1520)$ hyperon photoproduction from the nucleon:
\begin{equation}
\frac{d\sigma_{{\gamma}n\to K^0{\Lambda(1520)}}(\sqrt{s},\theta_{\Lambda^*}^*)}{d{\bf \Omega}_{\Lambda^*}^*}=
\frac{d\sigma_{{\gamma}p\rightarrow K^+\Lambda(1520)}({\sqrt{s}},{\theta_{\Lambda^*}^*})}{d{\bf \Omega}_{\Lambda^*}^*}.
\end{equation}
However, in the literature there is [33] an another option to choose this cross section. Thus, the authors of
[33], using the effective Lagrangian method, have obtained that the total and differential cross sections of the
${\gamma}n\rightarrow K^0\Lambda(1520)$ reaction are much smaller than those of the
${\gamma}p\rightarrow K^+\Lambda(1520)$ process
\footnote{It should be pointed out that this process has been theoretically studied also in the works [34--38].}
in the energy region $E_{\gamma} \le 3$~GeV
\footnote{Which is in line with the experimental findings of [28], where a strong suppression of the
$\Lambda(1520)$ production from neutrons compared to that from protons was observed at forward
$\Lambda(1520)$ angles in c.m.s. of photon and target nucleon.}
.
Therefore, to see the sensitivity of A dependence of the relative $\Lambda(1520)$ production cross sections
in photon-nucleus collisions of our main interest to the cross section for its creation on neutron target,
we will also ignore the latter one in our calculations.

     It is of further interest to get the feeling about the dependence of the maximal $t^+$ and
minimal $t^-$ values of $t$
\footnote{The $t^-$ value of $t$ is its magnitude at $\theta_{K}^*=180^{\circ}$.}
for the ${\gamma}p\rightarrow K^+\Lambda(1520)$ reaction when it takes place on a free target proton.
Using (14)--(16), one easily gets:
\begin{equation}
t^{\pm}=m_K^2-\frac{(s-m_p^2)}{2s}\left[s+m_K^2-M_{\Lambda^*}^2\mp\lambda(s,m_{K}^2,M_{\Lambda^*}^{2})\right]
\end{equation}
$$
=m_K^2-\frac{(s-m_p^2)(s+m_K^2-M_{\Lambda^*}^2)}{2s}\left(1\mp\sqrt{1-\frac{4m_K^2s}
{(s+m_K^2-M_{\Lambda^*}^2)^2}}\right).
$$
At threshold, $s=(m_K+M_{\Lambda^*})^2$, and $t^{\pm}$ are:
\begin{equation}
t^{\pm}=\frac{m_K(m_p^2-m_KM_{\Lambda^*}-M_{\Lambda^*}^2)}{(m_K+M_{\Lambda^*})}=-0.534~{\rm {GeV}^2}.
\end{equation}
Employing the formula (24), we can obtain that at $E_{\gamma}=2$~GeV of interest as well as in the case
when the free target proton is at rest $t^+=-0.21$~GeV$^2$, $t^-=-1.38$~GeV$^2$ and
$t^+-t^-=1.17$~GeV$^2$.
When the collision energy squared $s \gg (m_K+M_{\Lambda^*})^2$, we have:
\begin{equation}
t^{+} \approx m_K^2-\frac{(s-m_p^2)m_K^2}{s},
\end{equation}
\begin{equation}
t^{-} \approx m_K^2-\frac{(s-m_p^2)(s+m_K^2-M_{\Lambda^*}^2)}{s}.
\end{equation}
It is seen that with increasing energy $t^+$ approaches to zero and $t^-$ tends to $-\infty$.

   For the total $\Lambda(1520)$ in-medium width, $\Gamma_{\rm tot}$, appearing in (5)
and used in the subsequent calculations of $\Lambda(1520)$ resonance attenuation in ${\gamma}A$ interactions,
   we will adopt in the same manner as in [25] two different scenarios: {\bf i)} no in-medium
effects and, correspondingly, the scenario with the free $\Lambda(1520)$ width
\footnote{The reason why this width is included in the calculations is explained in [25].}
;
{\bf ii)} the sum of the free $\Lambda(1520)$ width and its collisional width of the type [22]
55($\rho_N/\rho_0$) MeV, where $\rho_N$ is the local nucleon density inside the nucleus.
As before in [25], we neglect
the momentum dependence of the in-medium $\Lambda(1520)$ width $\Gamma_{\rm tot}$, found in [22],
in our calculations.
This enables us to obtain an upper estimate of the strength of the respective double differential
cross sections and has an insignificant influence on the observables like
the A dependence of the ratio
between the $\Lambda(1520)$ production cross section in heavy nucleus and a light one (see below).

   Consider now the two-step $\Lambda(1520)$ production mechanism.

\section*{3. Secondary $\Lambda(1520)$ production process}

\hspace{1.5cm} At initial energy of 2 GeV of our interest the following two-step $\Lambda(1520)$
production process with
an intermediate pion may contribute to the $\Lambda(1520)$ creation in ${\gamma}A$ collisions:
\begin{equation}
\gamma+N_1 \to 2\pi+N,
\end{equation}
\begin{equation}
\pi+N_2 \to K+\Lambda(1520).
\end{equation}
At photon energies $E_{\gamma} \sim 2$~GeV the total cross section of the two-body channel
${\gamma}N \to {\pi}N$ is significantly less than that of the process (28) [39, 40] and, therefore, we
ignored it in our calculations. At these energies the four-body channel ${\gamma}N \to 3{\pi}N$ is
expected to play also a minor role in the $\Lambda(1520)$ production in ${\gamma}A$ interactions
since, on the one hand, its total cross section is less than that of the reaction (28) by a factor of
about 2.5 [40] and, on the other hand, the pions are produced in it with momenta at which the secondary
subprocess (29), opening at the free threshold momentum of 1.68 GeV/c,
is energetically suppressed. Thus, for example, our calculations show that at
$E_{\gamma}=2$~GeV the maximum allowable momentum of $\pi$ in the channel ${\gamma}N \to 3{\pi}N$
occuring on a free target nucleon being at rest is equal to 1.68~GeV/c. Which means that the pion momenta
accessible in this channel are subthreshold momenta for the $\Lambda(1520)$ secondary production process (29).

 Using the results given in [25], we get the
following expression for the $\Lambda(1520)$ production cross section for ${\gamma}A$ reactions at small
laboratory angles of interest from the channel (29):
\begin{equation}
\frac{d\sigma_{{\gamma}A\to {\Lambda(1520)}X}^{({\rm sec})}
({\bf p}_{\gamma})}
{d{\bf p}_{\Lambda^*}}=I_{V}^{({\rm sec})}[A]
\sum_{\pi'=\pi^+,\pi^0,\pi^-}\int\limits_{4\pi}d{\bf \Omega}_{\pi}
\int\limits_{p_{\pi}^{{\rm abs}}}^{p_{\pi}^{{\rm lim}}
(\vartheta_{\pi})}p_{\pi}^{2}dp_{\pi}
\times
\end{equation}
$$
\times
\left[\frac{Z}{A}\left<\frac{d\sigma_{{\gamma}p\to {\pi'}X}({\bf p}_{\gamma},
{\bf p}_{\pi})}{d{\bf p}_{\pi}}\right>+
\frac{N}{A}\left<\frac{d\sigma_{{\gamma}n\to {\pi'}X}({\bf p}_{\gamma},
{\bf p}_{\pi})}{d{\bf p}_{\pi}}\right>\right]
\times
$$
$$
\times
\left[\frac{Z}{A}\left<\frac{d\sigma_{{\pi'}p\to K\Lambda(1520)}({\bf p}_{\pi},
{\bf p}_{\Lambda^*})}{d{\bf p}_{\Lambda^*}}\right>+
\frac{N}{A}\left<\frac{d\sigma_{{\pi'}n\to K\Lambda(1520)}({\bf p}_{\pi},
{\bf p}_{\Lambda^*})}{d{\bf p}_{\Lambda^*}}\right>\right],
$$
where
\begin{equation}
I_{V}^{({\rm sec})}[A]=2{\pi}A^2\int\limits_{0}^{R}r_{\bot}dr_{\bot}
\int\limits_{-\sqrt{R^2-r_{\bot}^2}}^{\sqrt{R^2-r_{\bot}^2}}dz
\rho(\sqrt{r_{\bot}^2+z^2})
\int\limits_{0}^{\sqrt{R^2-r_{\bot}^2}-z}dl
\rho(\sqrt{r_{\bot}^2+(z+l)^2})
\times
\end{equation}
$$
\times
\exp{\left[-\sigma_{{\gamma}N}^{{\rm tot}}A\int\limits_{-\sqrt{R^2-r_{\bot}^2}}^{z}
\rho(\sqrt{r_{\bot}^2+x^2})dx
-\sigma_{{\pi}N}^{{\rm tot}}A\int\limits_{z}^{z+l}
\rho(\sqrt{r_{\bot}^2+x^2})dx\right]}
\times
$$
$$
\times
\exp{\left[-\int\limits_{z+l}^{\sqrt{R^2-r_{\bot}^2}}\frac{dx}
{\lambda_{\Lambda^*}(\sqrt{r_{\bot}^2+x^2},M_{\Lambda^*})}\right]},
$$
\begin{equation}
\left<\frac{d\sigma_{{\gamma}N\to {\pi'}X}({\bf p}_{\gamma},
{\bf p}_{\pi})}
{d{\bf p}_{\pi}}\right>=
\int\int
P({\bf p}_t,E)d{\bf p}_tdE
\left[\frac{d\sigma_{{\gamma}N\to {\pi'}X}(\sqrt{s},{\bf p}_{\pi})}
{d{\bf p}_{\pi}}\right],
\end{equation}
\begin{equation}
\left<\frac{d\sigma_{{\pi'}N\to K\Lambda(1520)}({\bf p}_{\pi},
{\bf p}_{\Lambda^*})}
{d{\bf p}_{\Lambda^*}}\right>=
\int\int
P({\bf p}_t,E)d{\bf p}_tdE
\left[\frac{d\sigma_{{\pi'}N\to K\Lambda(1520)}(\sqrt{s_1},{\bf p}_{\Lambda^*})}
{d{\bf p}_{\Lambda^*}}\right];
\end{equation}
\begin{equation}
  s_1=(E_{\pi}+E_{t})^2-(p_{\pi}{\bf \Omega_{\gamma}}+{\bf p}_{t})^2,
\end{equation}
\begin{equation}
 p_{\pi}^{{\rm lim}}(\vartheta_{\pi}) =
\frac{{\beta}_{A}p_{\gamma}\cos{\vartheta_{\pi}}+
 (E_{\gamma}+M_A)\sqrt{{\beta}_{A}^2-4m_{\pi}^{2}(s_{A}+
p_{\gamma}^{2}\sin^{2}{\vartheta_{\pi}})}}{2(s_{A}+
p_{\gamma}^{2}\sin^{2}{\vartheta_{\pi}})},
\end{equation}
\begin{equation}
 {\beta}_A=s_{A}+m_{\pi}^{2}-M_{A}^{2},\,\,s_A=(E_{\gamma}+M_A)^2-p_{\gamma}^{2},
\end{equation}
\begin{equation}
\cos{\vartheta_{\pi}}={\bf \Omega}_{\gamma}{\bf \Omega}_{\pi},\,\,\,\,
{\bf \Omega}_{\gamma}={\bf p}_{\gamma}/p_{\gamma},\,\,\,\,{\bf \Omega}_{\pi}={\bf p}_{\pi}/p_{\pi}.
\end{equation}
Here, $d\sigma_{{\gamma}N\to {\pi'}X}(\sqrt{s},{\bf p}_{\pi})/d{\bf p}_{\pi}$ are the
inclusive differential cross sections for pion production from
the primary photon-induced reaction channel (28), and the definition of other quantities
in (30)--(36) is identical to that in [25].
It should be noted that a distortion of the intermediate pion is described in eq. (31) by the
exponential factor in which the total but not inelastic ${\pi}N$ cross section is used.
This is due to the following. At beam energy of 2 GeV of interest the maximum allowable momentum
of a pion in the process ${\gamma}N \to 2{\pi}N$ taking place on a free resting nucleon amounts
to 1.85 GeV/c. This momentum is close to the threshold momentum of the channel
${\pi}N \to K\Lambda(1520)$. Therefore, quasielastic ${\pi}N$ scatterings, hindered by Pauli blocking,
will essentially suppress the pion ability to produce the $\Lambda(1520)$. This means that there is
an additional--to that caused by inelastic ${\pi}N$ interactions--loss of pion flux with respect to
the $\Lambda(1520)$ production due to these scatterings and its overall distortion should be
described by the total ${\pi}N$ cross section. For the relevant high energy pion, we can neglect
its distortion due to its absorption by two nucleons.

        In our method the differential cross sections
$d\sigma_{{\gamma}p\to {\pi'}X}(\sqrt{s},{\bf p}_{\pi})/d{\bf p}_{\pi}$ and
$d\sigma_{{\gamma}n\to {\pi'}X}(\sqrt{s},{\bf p}_{\pi})/d{\bf p}_{\pi}$ for pion production
in ${\gamma}p$ and ${\gamma}n$ collisions (28) have been described by the three-body phase space calculations
corrected for Pauli blocking (see, eq. (46) below)
leading to the suppression of the phase space available for the final-state nucleon
in these collisions.
In order to introduce this effect in our calculations, we have modified the nuclear spectral function
according to [41]:
\begin{equation}
P({\bf p}_t,E) \Longrightarrow P({\bf p}_t,E)\theta(|{\bf p}_{N}|-{\bar p}_F),
\end{equation}
where $\theta(x)$ stands for the step function, ${\bf p}_{N}$ is the momentum of the final-state nucleon
and ${\bar p}_F$ is the average Fermi momentum of the nucleus defined as [42]
\begin{equation}
{\bar p}_F=\left(\frac{3\pi^2<\rho_N>}{2}\right)^{1/3}.
\end{equation}
Here, $<\rho_N>$ is the average nucleon density. For $<\rho_N>=\rho_0/2$ [43] and $\rho_0=0.16$~fm$^{-3}$,
eq. (39) leads to ${\bar p}_F=210$~MeV/c
\footnote{It is interesting to note that this value of ${\bar p}_F$ is close to that of
${\bar p}_F=225$~MeV/c, obtained in [41] for a carbon target using the definition of ${\bar p}_F$
in the form ${\bar p}_F=\int d^3r{\rho_N({\bf r})}p_F({\bf r})/A$ with
$p_F({\bf r})=\left(3\pi^2\rho_N({\bf r})/2\right)^{1/3}$.}
.
We will use this magnitude throughout our calculations. The inclusion of Pauli blocking for final-state
nucleon results in a suppression of pion yields at high laboratory pion momenta, which in turn tends,
as we shall see later, to the quenching of the cross sections for $\Lambda(1520)$ production from
secondary channel ${\pi}N \to K\Lambda(1520)$ in the region of high $\Lambda(1520)$ momenta and
practically unaffects these cross sections at their "low" and intermediate momenta.

  Finally, using the results given in [44] as well as accounting for that the total cross sections
for the reactions ${\gamma}n \to {\pi}^{+}{\pi}^{-}n$ and ${\gamma}n \to {\pi}^{-}{\pi}^{0}p$ are
very similar, respectively, to those of the processes ${\gamma}p \to {\pi}^{+}{\pi}^{-}p$ and
${\gamma}p \to {\pi}^{+}{\pi}^{0}n$ [40, 45], one has:
\begin{equation}
\frac{d\sigma_{{\gamma}p\to {\pi^+}X}(\sqrt{s},{\bf p}_{\pi})}
{d{\bf p}_{\pi}}=\left[\sigma_{{\gamma}p \to {\pi}^{+}{\pi}^{-}p}(\sqrt{s})+
\sigma_{{\gamma}p \to {\pi}^{+}{\pi}^{0}n}(\sqrt{s})\right]f_3(s,{\bf p}_{\pi}),
\end{equation}
\begin{equation}
\frac{d\sigma_{{\gamma}p\to {\pi^0}X}(\sqrt{s},{\bf p}_{\pi})}
{d{\bf p}_{\pi}}=\sigma_{{\gamma}p \to {\pi}^{+}{\pi}^{0}n}(\sqrt{s})f_3(s,{\bf p}_{\pi}),
\end{equation}
\begin{equation}
\frac{d\sigma_{{\gamma}p\to {\pi^-}X}(\sqrt{s},{\bf p}_{\pi})}
{d{\bf p}_{\pi}}=\sigma_{{\gamma}p \to {\pi}^{+}{\pi}^{-}p}(\sqrt{s})f_3(s,{\bf p}_{\pi})
\end{equation}
and
\begin{equation}
\frac{d\sigma_{{\gamma}n\to {\pi^+}X}(\sqrt{s},{\bf p}_{\pi})}
{d{\bf p}_{\pi}}=\frac{d\sigma_{{\gamma}p\to {\pi^-}X}(\sqrt{s},{\bf p}_{\pi})}{d{\bf p}_{\pi}},
\end{equation}
\begin{equation}
\frac{d\sigma_{{\gamma}n\to {\pi^0}X}(\sqrt{s},{\bf p}_{\pi})}
{d{\bf p}_{\pi}}=\frac{d\sigma_{{\gamma}p\to {\pi^0}X}(\sqrt{s},{\bf p}_{\pi})}{d{\bf p}_{\pi}},
\end{equation}
\begin{equation}
\frac{d\sigma_{{\gamma}n\to {\pi^-}X}(\sqrt{s},{\bf p}_{\pi})}
{d{\bf p}_{\pi}}=\frac{d\sigma_{{\gamma}p\to {\pi^+}X}(\sqrt{s},{\bf p}_{\pi})}{d{\bf p}_{\pi}},
\end{equation}
where
\begin{equation}
f_3(s,{\bf p}_{\pi})
=\frac{{\pi}}{4I_{3}(s,m_N,m_{\pi},m_{\pi})E_{\pi}}
\frac{\lambda(s_{N\pi},m_{N}^{2},m_{\pi}^{2})}{s_{N\pi}}F_{{\rm block}},
\end{equation}
\begin{equation}
I_{3}(s,m_N,m_{\pi},m_{\pi})=(\frac{{\pi}}{2})^2
\int\limits_{(m_{N}+m_{\pi})^2}^{({\sqrt{s}}-m_{\pi})^2}
\frac{\lambda(s_{N\pi},m_{N}^{2},m_{\pi}^{2})}{s_{N\pi}}
\frac{\lambda(s,s_{N\pi},m_{\pi}^{2})}{s}\,ds_{N\pi},
\end{equation}
\begin{equation}
s_{N\pi}=s+m_{\pi}^{2}-2(E_{\gamma}+E_t)E_{\pi}+
2({\bf p}_{\gamma}+{\bf p}_t){\bf p}_{\pi}.
\end{equation}
Here, $\sigma_{{\gamma}p \to {\pi}^{+}{\pi}^{-}p}(\sqrt{s})$ and
$\sigma_{{\gamma}p \to {\pi}^{+}{\pi}^{0}n}(\sqrt{s})$
are the free total cross sections of the reactions ${\gamma}p \to {\pi}^{+}{\pi}^{-}p$ and
${\gamma}p \to {\pi}^{+}{\pi}^{0}n$ calculated at the "off-shell" collision energy $\sqrt{s}$,
$m_{\pi}$ is the rest pion mass and $F_{\rm block}$ is the Pauli blocking factor, which is given by:.
\begin{equation}
F_{\rm block}=
\left\{
\begin{array}{lll}
	1
	&\mbox{for ${\bar E}_{F} \le E_{N}^{-}$}, \\
	&\\
        \frac{(E_N^+-{\bar E}_F)}{(E_N^+-E_N^-)}
	&\mbox{for $E_{N}^{-}< {\bar E}_{F} \le E_N^+$}, \\
                   &\\
                   0
                   &\mbox{for ${\bar E}_{F} > E_N^+$},
\end{array}
\right.	
\end{equation}
where
\begin{equation}
E_N^{\pm}=\frac{{\tilde {\omega}}{\beta}\pm{\tilde {Q}}\lambda(s_{N\pi},m_{N}^{2},m_{\pi}^{2})}{2s_{N\pi}}
\end{equation}
and
\begin{equation}
{\tilde {\omega}}=E_{\gamma}+E_t-E_{\pi},\,\,\,\,\,{\tilde Q}=|{\bf p}_{\gamma}+{\bf p}_{t}-{\bf p}_{\pi}|,
\end{equation}
\begin{equation}
\beta=s_{N\pi}+m_N^2-m_{\pi}^2,\,\,\,\,\,\,{\bar E}_{F}=\sqrt{{\bar p}_{F}^2+m_{N}^2}.
\end{equation}

  In our calculations of the cross sections for $\Lambda(1520)$ production from secondary channel (29) in
${\gamma}A$ interactions we have neglected the contribution to this production from elementary process
${\gamma}N \to 2{\pi^0}N$, since its cross section is expected to be low [40, 46] at photon energy of our
interest. For the free total cross sections $\sigma_{{\gamma}p \to \pi^+\pi^-p}$ and
 $\sigma_{{\gamma}p \to \pi^+\pi^0n}$, entering into eqs. (40)--(42), we used the following fits of the
available [40, 47] experimental and theoretical information on them in energy region
$1~{\rm GeV} < E_{\gamma} < 4~{\rm GeV}$:
\begin{equation}
\sigma_{{\gamma}p \to \pi^+\pi^-p}(\sqrt{s})=74\left(E_{\gamma}/GeV\right)^{-0.90}~[{\rm {{\mu}b}}],
\end{equation}
\begin{equation}
\sigma_{{\gamma}p \to \pi^+\pi^0n}(\sqrt{s})=53\left(E_{\gamma}/GeV\right)^{-0.85}~[{\rm {{\mu}b}}],
\end{equation}
in which in our case the initial photon energy $E_{\gamma}$ should be expressed via the collision
energy $\sqrt{s}$ through the use of eq. (22).

The elementary $\Lambda(1520)$ production reactions  ${\pi}^+n \to K^+\Lambda(1520)$,
${\pi}^0p \to K^+\Lambda(1520)$, ${\pi}^0n \to K^0\Lambda(1520)$ and
${\pi}^-p \to K^0\Lambda(1520)$ have been included in our calculations of the $\Lambda(1520)$ production on nuclei.
The differential cross sections of these reactions were completely taken from [25].

  Now, we discuss the results of our calculations for $\Lambda(1520)$
  production in ${\gamma}A$ interactions within the model outlined above.

\section*{4. Results}

\hspace{1.5cm} Let us take up first the absolute cross sections for $\Lambda(1520)$ production
in ${\gamma}A$ ($A=$$^{12}$C, $^{63}$Cu, and $^{197}$Au) collisions calculated
for $E_{\gamma}=2$ GeV and for the laboratory $\Lambda(1520)$
production angle of 0$^{\circ}$. The cross sections obtained using the free $\Lambda(1520)$ width
for the total $\Lambda(1520)$ in-medium width are shown in figure 1,
\begin{figure}[htb]
{\centering
\includegraphics[width=5.8cm]{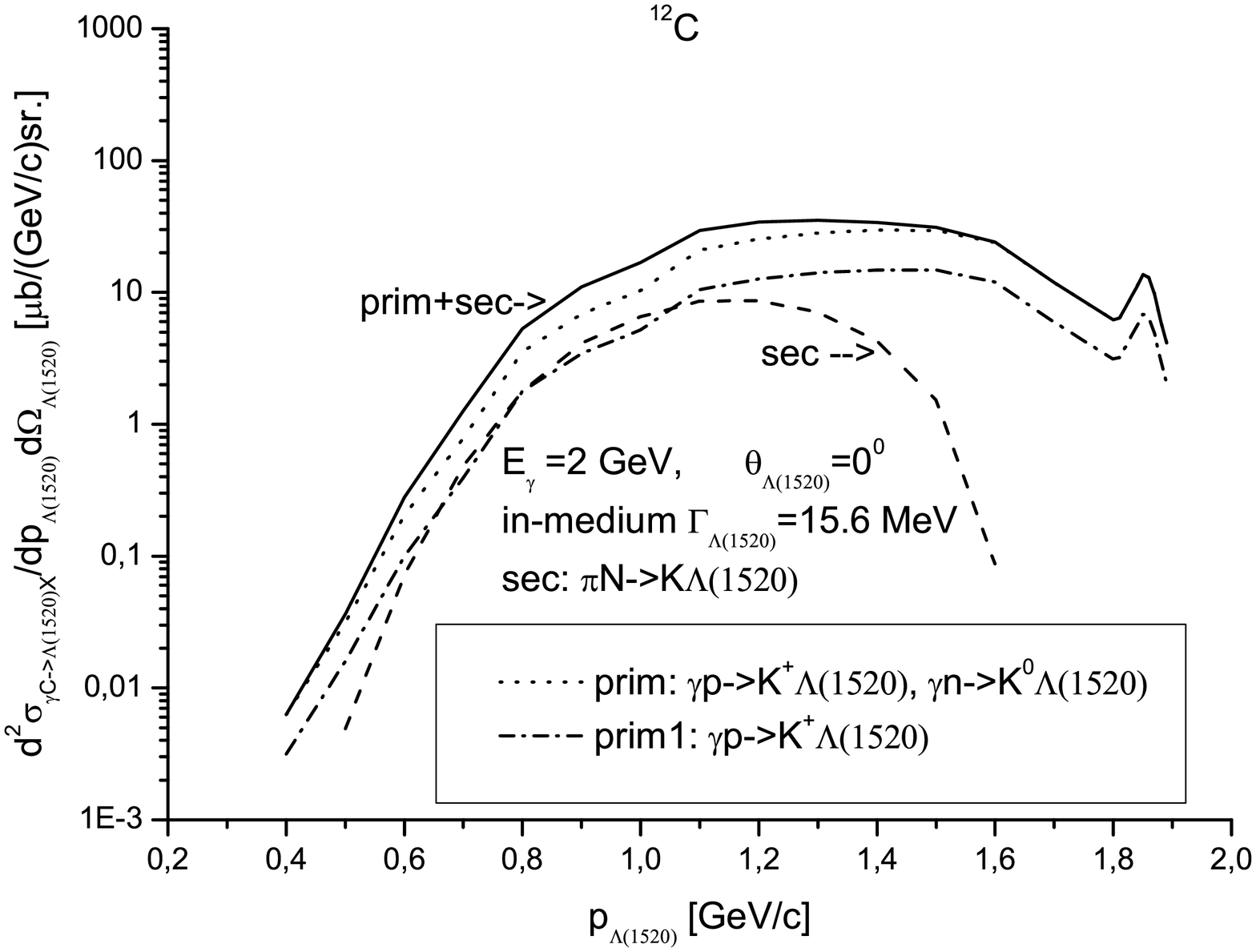}
\hfill
\includegraphics[width=5.8cm]{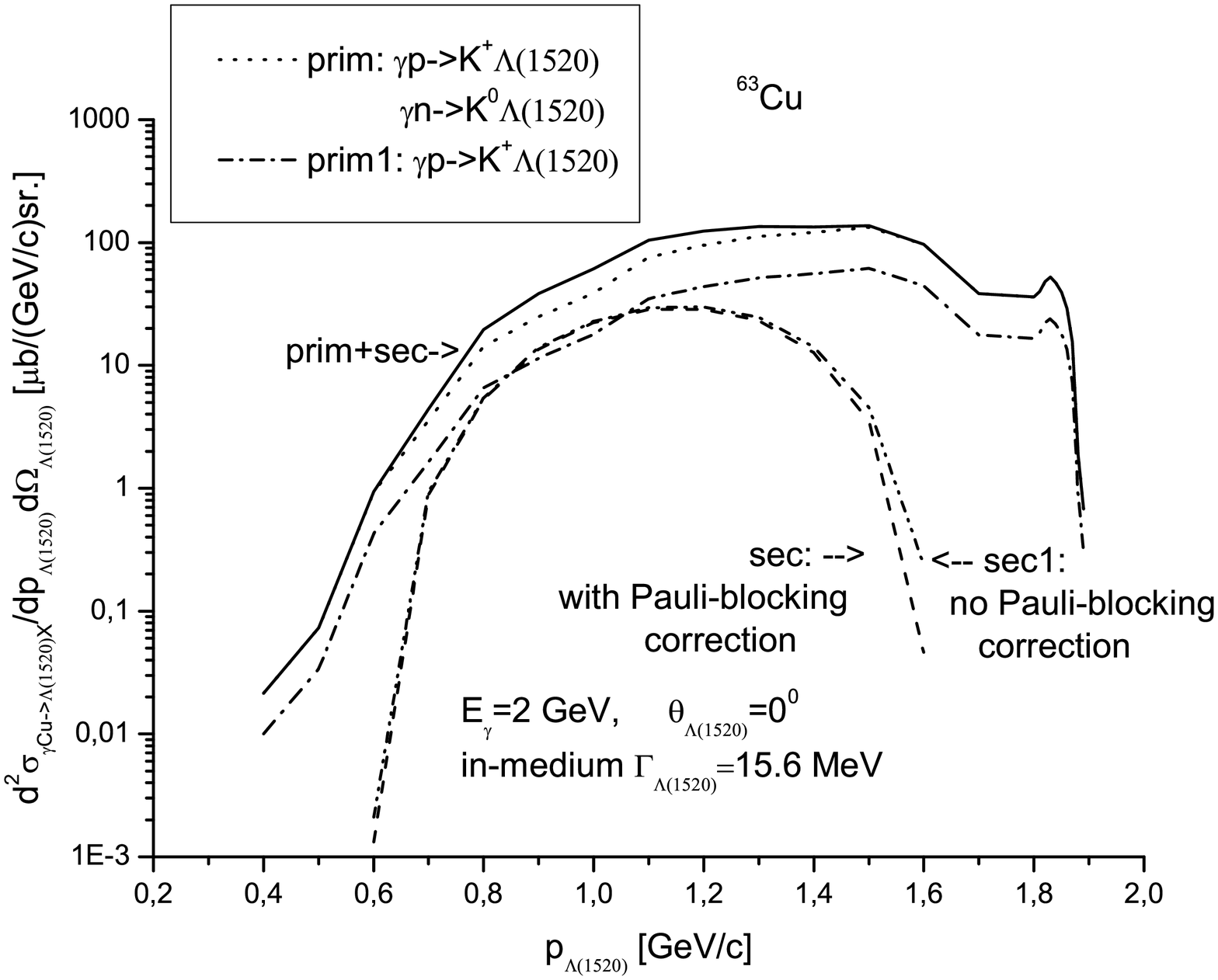}
\hfill
\includegraphics[width=5.8cm]{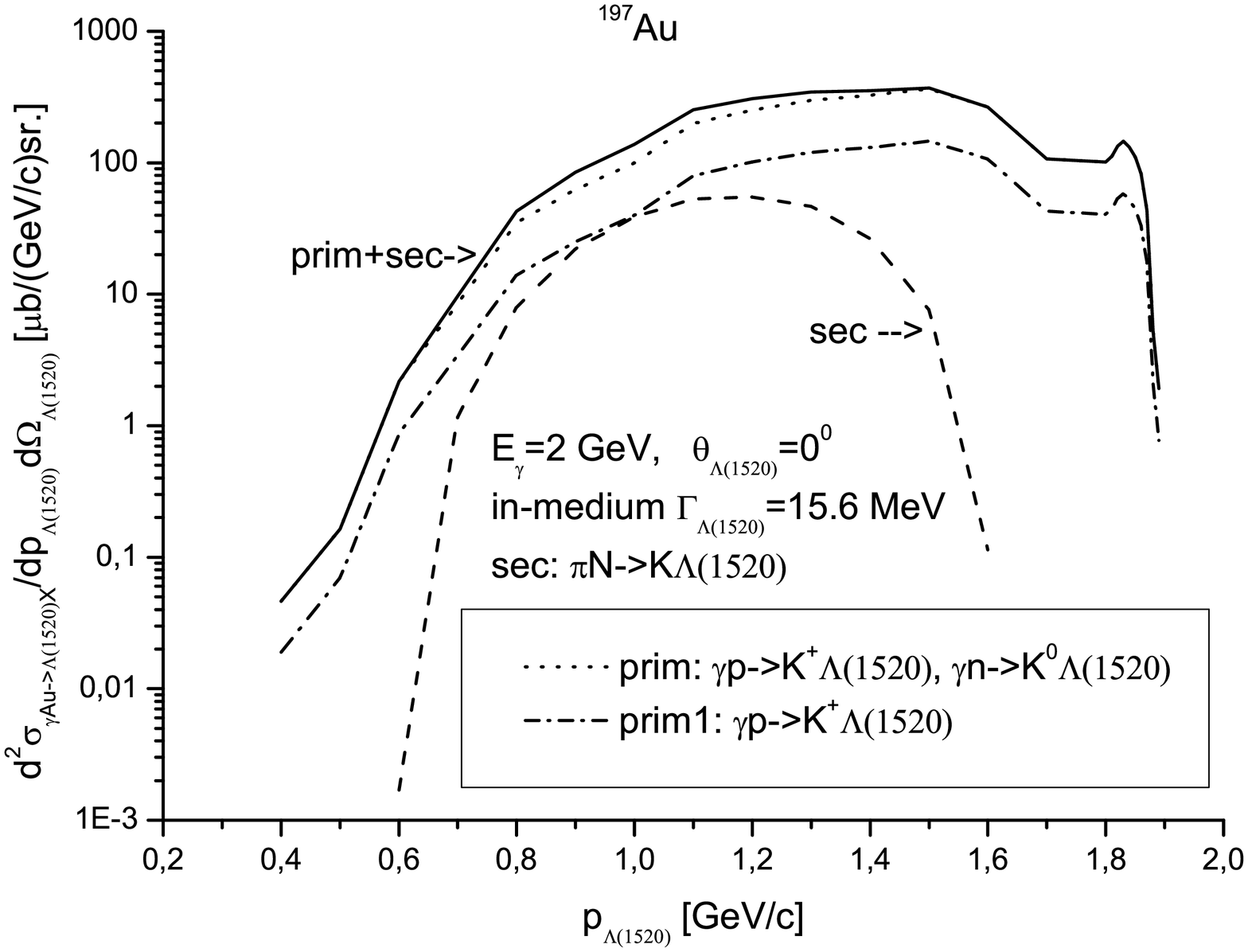}}
\vspace*{-2mm} \caption{Double differential cross sections for the production of $\Lambda(1520)$ hyperons
on $^{12}$C (left panel), $^{63}$Cu (middle panel) and $^{197}$Au (right panel) nuclei
at $E_{\gamma}=2$ GeV and a lab angle of 0$^{\circ}$ as
functions of $\Lambda(1520)$ momentum. The dotted and dash-dotted lines are calculations for the
one-step $\Lambda(1520)$ creation mechanism assuming, respectively, the isospin symmetry for the
$\Lambda(1520)$ hyperon photoproduction from nucleon and its photoproduction only on the proton target
in the process ${\gamma}p \to K^+\Lambda(1520)$.
The dashed and two dot-dashed lines are calculations for the two-step $\Lambda(1520)$ production
mechanism, respectively, with and without Pauli-blocking correction for the final-state nucleon
in the process ${\gamma}N \to 2{\pi}N$.
The solid line is the sum of the dotted and dashed lines. The
loss of $\Lambda(1520)$ hyperons in nuclear matter was determined by their free width.}
\label{void}
\end{figure}
and the ones calculated for the modified $\Lambda(1520)$ width in the medium
are given in figure 2.
\begin{figure}[htb]
{\centering
\includegraphics[width=5.8cm]{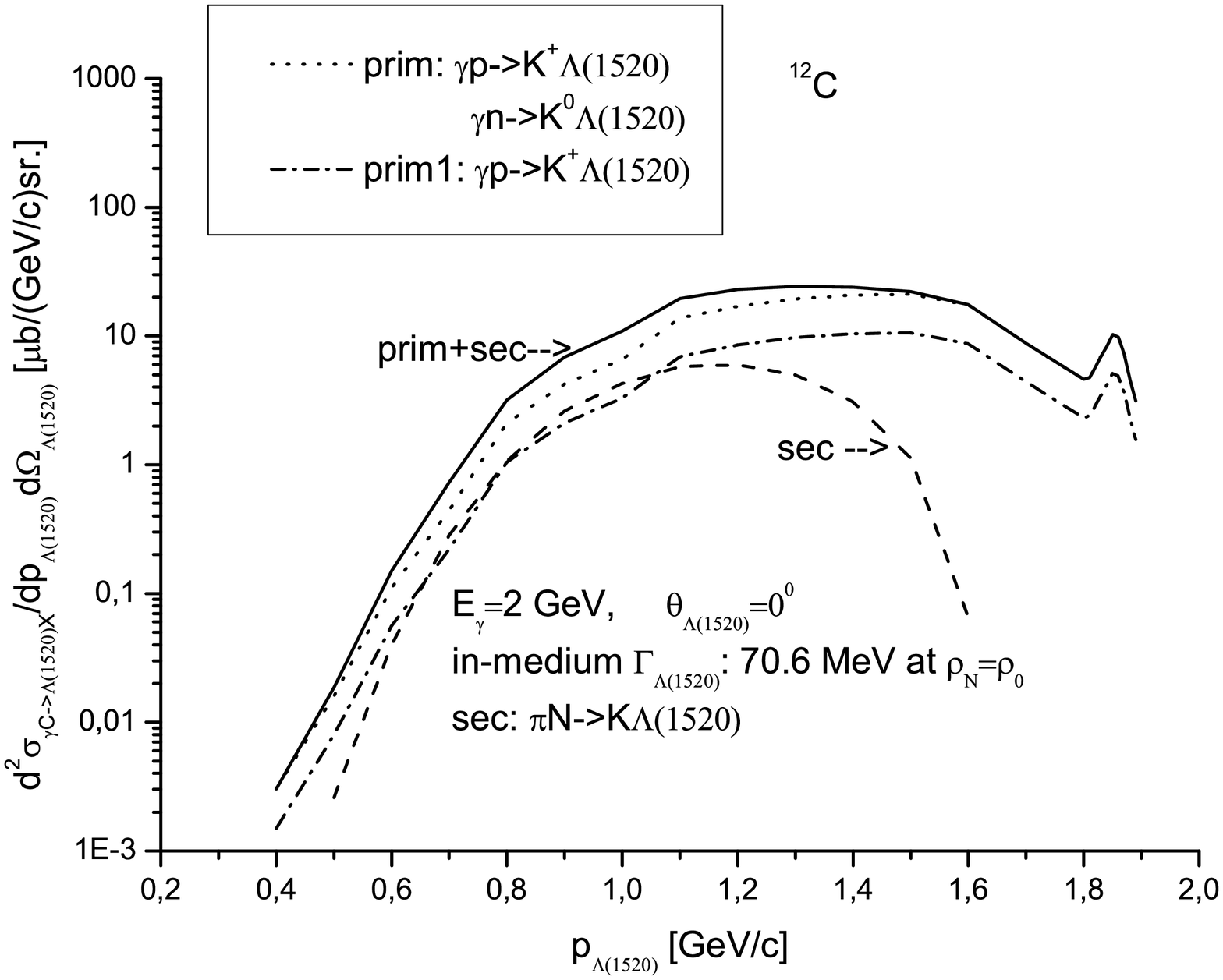}
\hfill
\includegraphics[width=5.8cm]{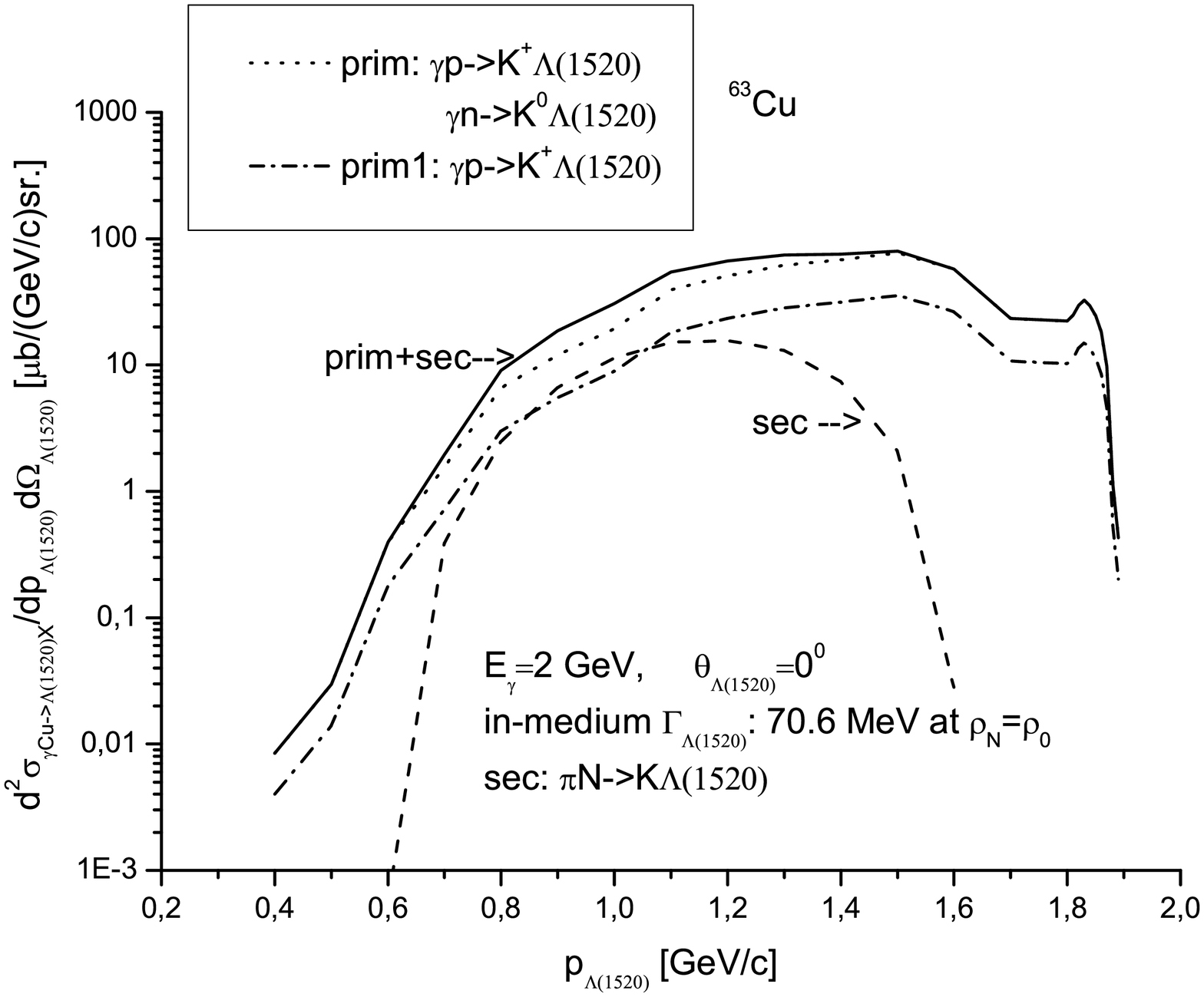}
\hfill
\includegraphics[width=5.8cm]{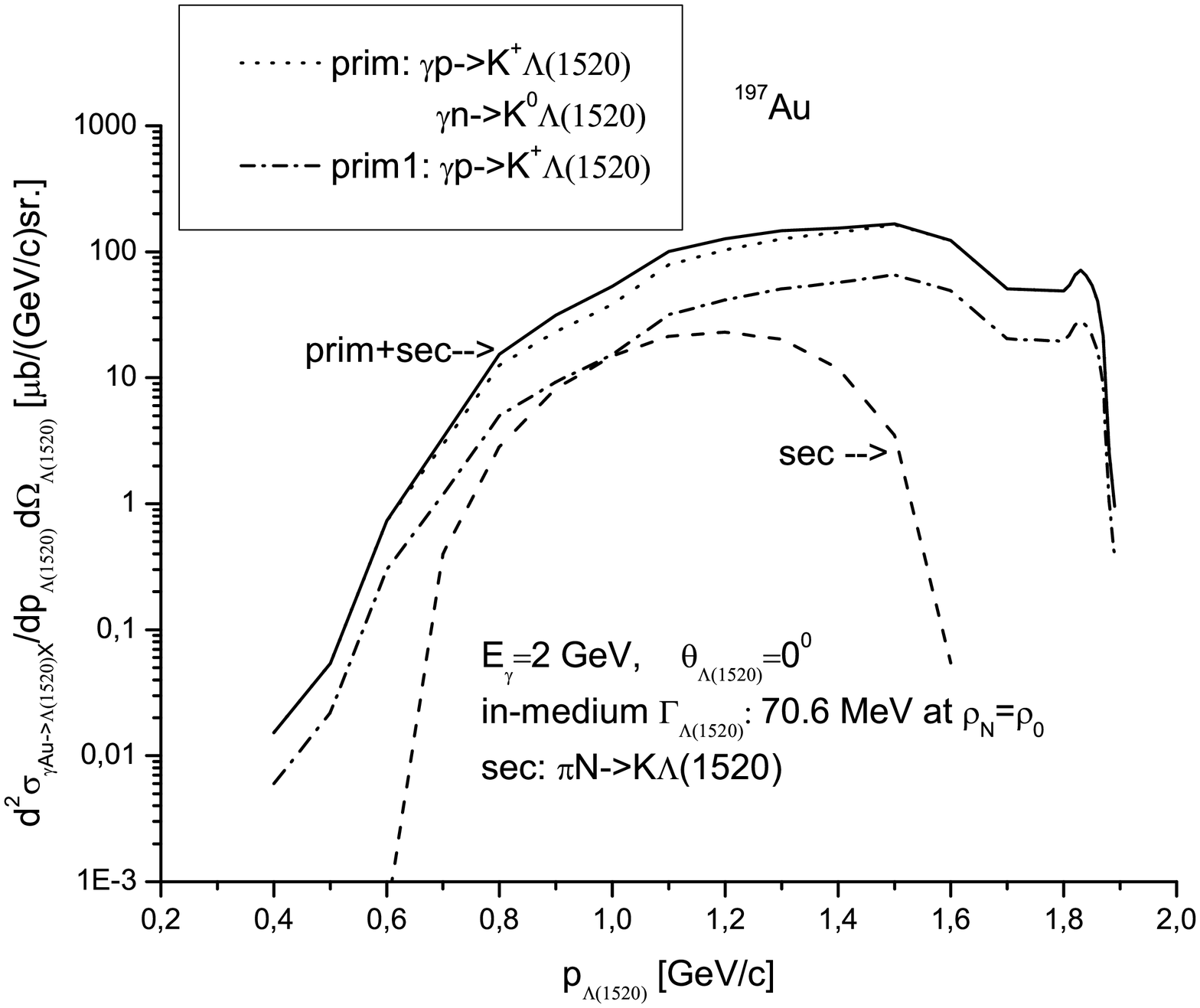}}
\vspace*{-2mm} \caption{Double differential cross sections for the production of $\Lambda(1520)$ hyperons
on $^{12}$C (left panel), $^{63}$Cu (middle panel) and $^{197}$Au (right panel) nuclei
at $E_{\gamma}=2$ GeV and a lab angle of 0$^{\circ}$ as functions of $\Lambda(1520)$ momentum.
The notation of the curves is essentially similar to that in figure 1,
with the only difference that the
absorption of $\Lambda(1520)$ hyperons in nuclei was governed by their modified in the medium width.}
\label{void}
\end{figure}
A close look at these figures will show that the one-step $\Lambda(1520)$ production mechanism
plays the dominant role
at $\Lambda(1520)$ momenta $\le$ 0.8 GeV/c and $\ge$ 1.2 GeV/c
for all considered target nuclei as well as for both adopted scenarios for the
$\Lambda(1520)$ in-medium width and for the $\Lambda(1520)$ photoproduction on the neutron,
whereas at intermediate $\Lambda(1520)$ momenta
$0.8~{\rm {GeV/c}} \le p_{\Lambda(1520)} \le 1.2~{\rm {GeV/c}}$
its dominance is less pronounced when the isospin symmetry for the $\Lambda(1520)$ photoproduction
from nucleon is assumed to be. In the case when the $\Lambda(1520)$ photoproduction takes place only
on the proton target in the primary process ${\gamma}p \to K^+\Lambda(1520)$, the contributions from
the one-step and two-step $\Lambda(1520)$ creation mechanisms are comparable here.
This means
that the channel ${\pi}N \to K\Lambda(1520)$ has to be taken into consideration on close examination
of the A dependence of the relative $\Lambda(1520)$ hyperon production cross section in photon--nucleus reactions
at energies just above threshold with the aim of extracting of the information on the $\Lambda(1520)$ width in
nuclear medium. Comparing the results of our full calculations (the sum of contributions both from primary and from
secondary $\Lambda(1520)$ production processes) presented in figures 1 and 2 by solid lines, we
see yet that for given target nucleus there is clear difference between the results obtained by using different
$\Lambda(1520)$ in-medium widths under consideration.
We may see, for example, that for the considered $^{12}$C, $^{63}$Cu and $^{197}$Au nuclei
the cross section in the momentum range $\sim$ 1.2--1.6 GeV/c (where it is the greatest)
when calculated with the increased $\Lambda(1520)$ width in the medium
is reduced, respectively, by a factors of about 1.4, 1.8 and 2.3 compared to that obtained in the case
when the loss of $\Lambda(1520)$ hyperons in nuclear matter was determined by their free width.
The inclusion of Pauli blocking for final nucleon participating in the process
${\gamma}N \to 2{\pi}N$, as may be seen from figure 1, while leaving unaffected the $\Lambda(1520)$
production cross section from secondary channel ${\pi}N \to K\Lambda(1520)$ at momenta $\le$ 1.4 GeV/c,
leads to an appreciable quenching of this cross section in the region of higher $\Lambda(1520)$ momenta.
We will account for this Pauli blocking throughout our calculations.

     An inspection of figure 3, where the double differential cross sections for the production of
$\Lambda(1520)$ hyperons on $^{12}$C and $^{197}$Au target nuclei from the one- plus two-step
$\Lambda(1520)$ creation processes given before separately in figures 1 and 2  by solid lines
are presented together to see more clearly their sensitivity to the choice of the
$\Lambda(1520)$ width in the medium,
shows that for heavy nuclei like $^{197}$Au there are indeed experimentally observed
changes in these cross sections due to the modified in the medium $\Lambda(1520)$ width.
\begin{figure}[htb]
\begin{center}
\includegraphics[width=9.0cm]{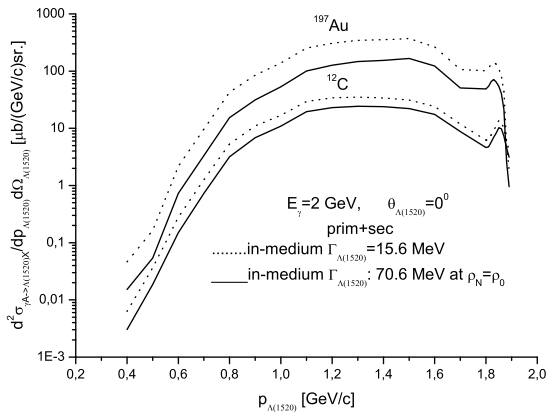}
\vspace*{-2mm} \caption{Double differential cross sections for the production of $\Lambda(1520)$ hyperons
on $^{12}$C (two lower lines) and $^{197}$Au (two upper lines) target nuclei
at $E_{\gamma}=2$ GeV and an angle of 0$^{\circ}$
as functions of $\Lambda(1520)$ momentum for the primary plus secondary $\Lambda(1520)$
creation processes. The dotted lines are the calculations with the free $\Lambda(1520)$ width.
The solid lines are the calculations with the modified $\Lambda(1520)$ width in the medium.}
\label{void}
\end{center}
\end{figure}

Now that we have presented the absolute $\Lambda(1520)$ production cross sections from ${\gamma}A$
reactions, we need to focus upon the following relative observable--the ratio
$R(^{A}X)/R(^{12}{\rm C})=
({\tilde \sigma}_{{\gamma}A}(p_{\Lambda^*},0^{\circ})/A)/ \\
({\tilde \sigma}_{{\gamma}^{12}{\rm C}}(p_{\Lambda^*},0^{\circ})/12)$, where
${\tilde \sigma}_{{\gamma}A}(p_{\Lambda^*},0^{\circ})$ is
the double differential cross section for the production
of $\Lambda(1520)$ hyperons with momentum $p_{\Lambda^*}$ at a lab angle of 0$^{\circ}$ in
these reactions. Along with analogous ratio in $pA$ collisions [25], it can be considered also as a
measure for the $\Lambda(1520)$ width in nuclear matter in ${\gamma}A$ interactions.
\begin{figure}[htb]
{\centering
\includegraphics[width=9.0cm]{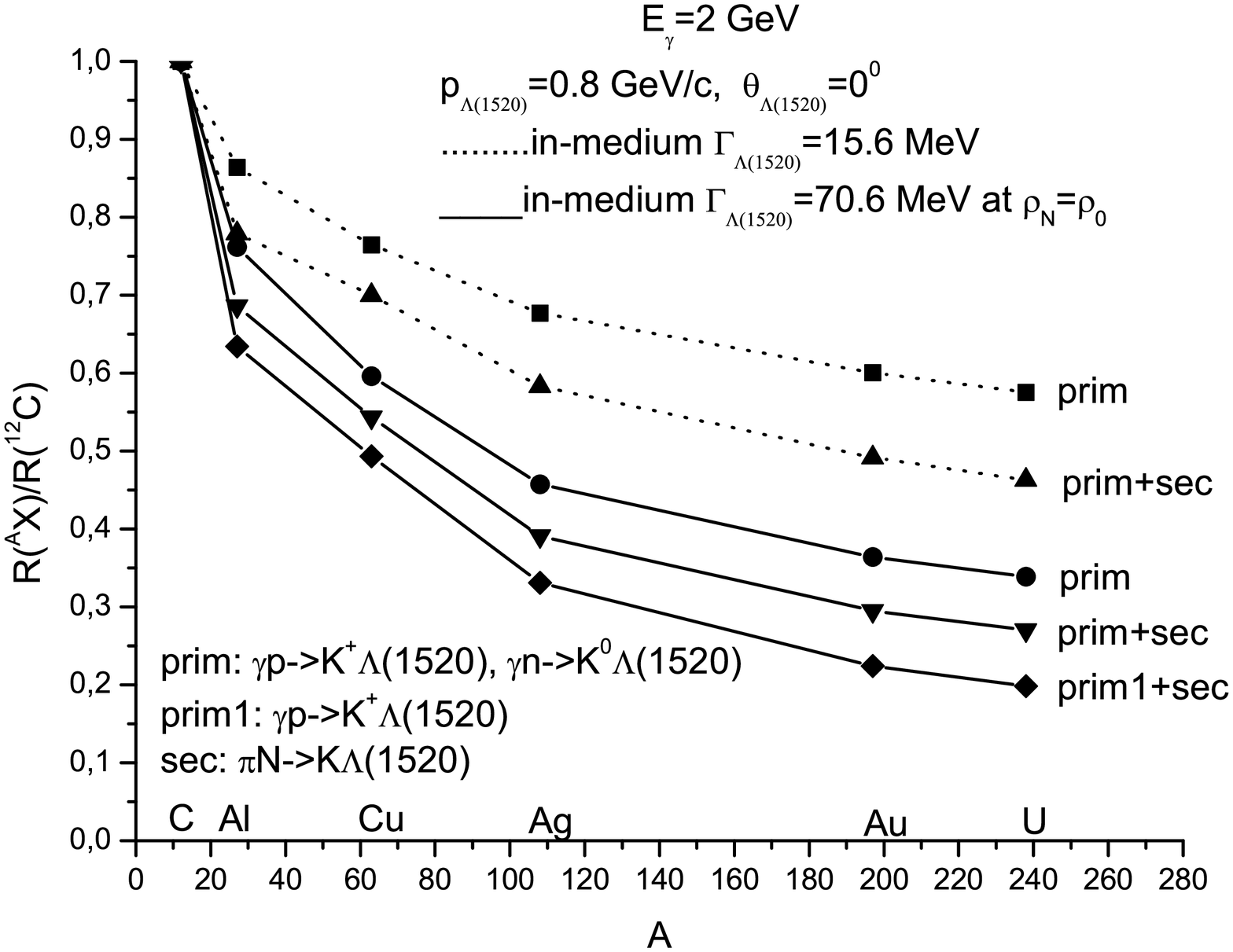}
\hfill
\includegraphics[width=9.0cm]{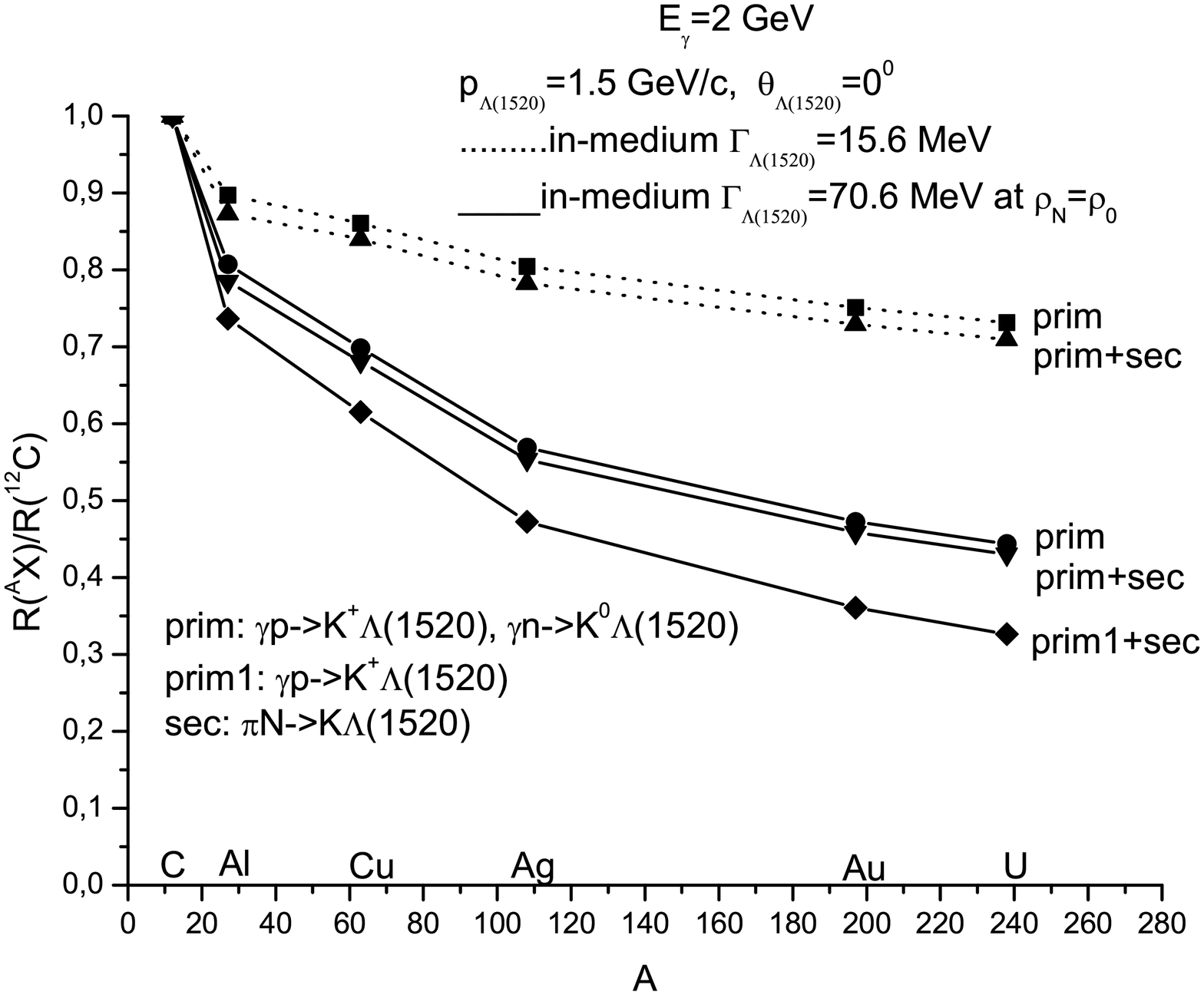}}
\vspace*{-2mm} \caption{A dependence of the ratio $R(^{A}X)/R(^{12}{\rm C})$
at $E_{\gamma}=2$ GeV as well as for the
$\Lambda(1520)$ momenta of 0.8 GeV/c (left panel) and 1.5 GeV/c (right panel).
The loss of $\Lambda(1520)$ hyperons in nuclear matter was governed by their free width
(dotted lines) and by their modified width (solid lines). For the rest of notation see the text.}
\label{void}
\end{figure}

      Figure 4 shows the A dependence of the above ratio calculated for
$^{27}$Al, $^{63}$Cu, $^{108}$Ag, $^{197}$Au, $^{238}$U target nuclei
for the one-step and
one- plus two-step $\Lambda(1520)$ creation mechanisms (corresponding lines with symbols 'prim'
and 'prim+sec', 'prim1+sec' by them
\footnote{We remind that the symbols 'prim' and 'prim1' here denote, as before, the calculations
in which, respectively, the isospin symmetry was assumed for the $\Lambda(1520)$
photoproduction from the nucleon and was supposed that the $\Lambda(1520)$ photoproduction takes place
only on the proton target in the direct process ${\gamma}p \to K^+\Lambda(1520)$.}
)
at $E_{\gamma}=2$ GeV as well as for the $\Lambda(1520)$ momenta of 0.8 GeV/c, 1.5 GeV/c and for
two adopted options for the $\Lambda(1520)$ in-medium width.
It is seen from this figure that there are a measurable changes in the considered ratio
of the order of 20\% and 60\%, respectively, for middle and heavy target nuclei
due to the modified $\Lambda(1520)$ in-medium width (compare respective dotted and solid lines
in figure 4).
The secondary production process ${\pi}N \to K\Lambda(1520)$ influences insignificantly the ratio
at $\Lambda(1520)$ momentum of 1.5 GeV/c and has a strong effect on it
at $\Lambda(1520)$ momentum of 0.8 GeV/c (cf. figures 1 and 2).
At this momentum for heavy nuclei like Au, U the calculated ratio
$R(^{A}X)/R(^{12}{\rm C})$ can be of the order of 0.35 for the direct (prim) $\Lambda(1520)$ production mechanism
and 0.27 for the direct plus two-step (prim+sec) $\Lambda(1520)$ creation mechanisms for the modified
$\Lambda(1520)$ width in nuclear matter.
It is seen yet that the difference between the calculations with adopting the two considered scenarios
for the $\Lambda(1520)$ photoproduction on the neutron (between solid lines with symbols 'prim+sec' and
'prim1+sec' in figure 4) is of the order of 10\% and 30\% for middle and heavy target nuclei, respectively,
for both chosen Lambda momenta. This means that, although it is less by half than that between the results
obtained by using two different options for the $\Lambda(1520)$ in-medium width,
to put more strong constraints on this width from the observation of the A dependence like that just considered,
one needs at first to get the definite experimental information on the $\Lambda(1520)$ photoproduction
on the neutron target. It is interesting to note that, even though the secondary channel
${\pi}N \to K\Lambda(1520)$ contributes more largely to the absolute $\Lambda(1520)$ creation
cross section in heavier nuclei than in light ones, its inclusion leads to the reduction of the
transparency ratio of interest compared to that obtained by considering only primary photon--nucleon
$\Lambda(1520)$ production processes. This is due to the fact that in the chosen kinematics
the two-step to one-step $\Lambda(1520)$ production cross section ratio for ${\gamma}{\rm C}$ reactions
is greater than that for ${\gamma}A$ interactions.
\begin{figure}[htb]
\begin{center}
\includegraphics[width=9.0cm]{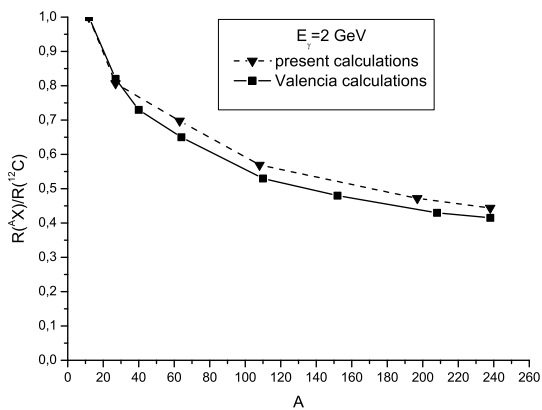}
\vspace*{-2mm} \caption{Ratio
$R(^{A}X)/R(^{12}{\rm C})$
as a function of the nuclear mass number
calculated for incident energy of 2 GeV
within the Valencia approach [24] (solid curve) and in the framework of the present model
(dashed line). For notation see the text.}
\label{void}
\end{center}
\end{figure}

As figure 2 shows and due to the kinematics,
the main contribution to the nuclear total $\Lambda(1520)$ production cross section from direct mechanism
comes from the $\Lambda(1520)$ momenta around momentum of 1.5 GeV/c and from small
$\Lambda(1520)$ production angles. Therefore, the ratio
$R(^{A}X)/R(^{12}{\rm C})=
({\tilde \sigma}_{{\gamma}A}(p_{\Lambda^*},0^{\circ})/Z)/
({\tilde \sigma}_{{\gamma}^{12}{\rm C}}(p_{\Lambda^*},0^{\circ})/6)$
calculated for the primary process ${\gamma}p \to K^+\Lambda(1520)$
at momentum of 1.5 GeV/c and for the employed momentum-independent
$\Lambda(1520)$ in-medium width can be compared to a first approximation
with the corresponding ratio of the nuclear cross sections obtained in [24] also at photon energy of 2 GeV
and for the nominal $\Lambda(1520)$ momentum-dependent in-medium width
(with the solid line in the left panel of figure 4 in [24]).
This comparison is shown in figure 5. By looking at this figure, we see that for targets heavier than the Al
target the difference between the two model calculations is insignificant (within a 5\%), which means that
the momentum dependence of the $\Lambda(1520)$ in-medium width adopted in [24] has a weak effect on the
relative observable under consideration.
\begin{figure}[htb]
\begin{center}
\includegraphics[width=9.0cm]{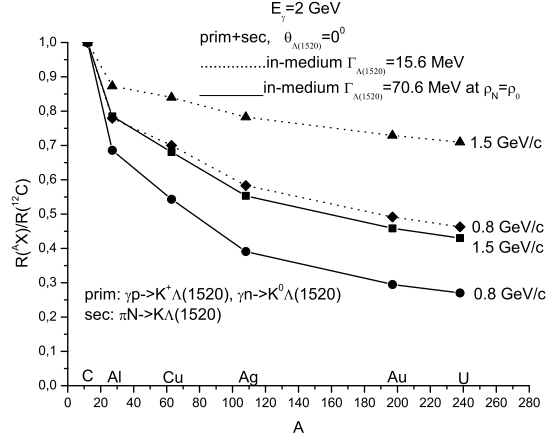}
\vspace*{-2mm} \caption{A dependence of the ratio $R(^{A}X)/R(^{12}{\rm C})$
for the one- plus two-step $\Lambda(1520)$ production mechanisms
as well as for the $\Lambda(1520)$ momenta of 0.8 GeV/c and 1.5 GeV/c. The dotted and solid
lines denote the same as in figure 4.}
\label{void}
\end{center}
\end{figure}
\begin{figure}[htb]
\begin{center}
\includegraphics[width=9.0cm]{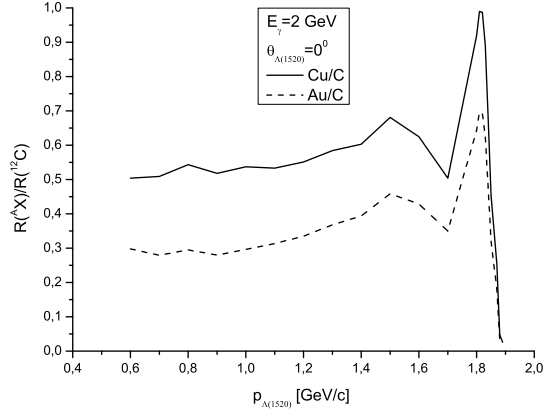}
\vspace*{-2mm} \caption{Ratio
$R(^{A}X)/R(^{12}{\rm C})=
({\tilde \sigma}_{{\gamma}A}(p_{\Lambda^*},0^{\circ})/A)/
({\tilde \sigma}_{{\gamma}^{12}{\rm C}}(p_{\Lambda^*},0^{\circ})/12)$
as a function of the $\Lambda(1520)$ momentum
for the one- plus two-step $\Lambda(1520)$ production mechanisms
calculated for incident energy of 2 GeV for $A=$Cu (solid line) and $A=$Au (dashed curve) as well as
for the modified in the medium $\Lambda(1520)$ width.
The isospin symmetry was assumed for the $\Lambda(1520)$ photoproduction from the nucleon.}
\label{void}
\end{center}
\end{figure}

  To see the sensitivity of the ratio $R(^{A}X)/R(^{12}{\rm C})$ of interest to the $\Lambda(1520)$
  momentum, in figure 6 we show together the results of our overall calculations for this ratio,
  given before separately in figure 4.
One can see that the differences between the calculations for the same $\Lambda(1520)$ width
(between two dotted and between two solid lines) are significant,
which means that this relative
observable depends strongly on the $\Lambda(1520)$ momentum in the momentum region
$0.8~{\rm {GeV/c}} \le p_{\Lambda(1520)} \le 1.5~{\rm {GeV/c}}$, where the cross section for $\Lambda(1520)$
production is sufficiently large.
This finding is in line with that of figure 7, where the detailed momentum dependence of the considered
ratio $R(^{A}X)/R(^{12}{\rm C})$ for $A=$Cu and Au in the studied momentum region is shown.
It is clearly seen that for both target nuclei this ratio increases with $\Lambda(1520)$ momentum
in the range of 0.9--1.5 GeV/c. At larger momenta it demonstrates more complicated behavior.
The measurement of such dependence should, in particular, reflect the fact that the $\Lambda(1520)$ width in the medium is momentum independent. Comparing the presented in figures 6 and 7 results, we may see that the
momentum dependence of the ratio $R(^{A}X)/R(^{12}{\rm C})$ under consideration is appreciably sensitive also
to the $\Lambda(1520)$ in-medium width.

   Taking into account the above considerations, we come to the conclusion that
the A and momentum dependences of
the absolute and relative cross sections for $\Lambda(1520)$ production in ${\gamma}A$ reactions
can be useful to help determine the $\Lambda(1520)$ width in cold nuclear matter.
Extracting this width from such dependences requires comparison with model calculations, in which
it is necessary to account for the secondary pion-induced $\Lambda(1520)$ production processes.

\section*{5. Summary}

\hspace{1.5cm} In this paper we have investigated the attenuation of $\Lambda(1520)$ hyperons in nuclei
in photon-induced reactions near the threshold on the basis of a collision model, which accounts for both
primary photon--nucleon and secondary pion--nucleon $\Lambda(1520)$ production processes.
We have discussed the mass and momentum dependences of the forward $\Lambda(1520)$ production in these
reactions at initial photon energy of 2 GeV. They were calculated for two options for the
$\Lambda(1520)$ in-medium width. We have shown that there are a measurable changes in these dependences
due to the modified in the medium width of the $\Lambda(1520)$. We have found also that the secondary
channel ${\pi}N \to K{\Lambda(1520)}$ materially affects the intermediate momentum $\Lambda(1520)$
production on nuclei. Consequently, it has to be taken into account in extracting the width of
$\Lambda(1520)$ hyperons in nuclear matter from their observed yields in ${\gamma}A$ interactions.
\\
\\


\begin{thebibliography}{99}
\bibitem{1} T. Ishikawa {\it et al.}, Phys. Lett. B {\bf 608}, 215 (2005).
\bibitem{2} C. Djalali {\it et al.}, J. Phys. G. {\bf 35}, 104035 (2008).
\bibitem{3} R. Nasseripour {\it et al.}, Phys. Rev. Lett. {\bf 99}, 262302 (2007).
\bibitem{4} M. H. Wood {\it et al.}, Phys. Rev. C {\bf 78}, 015201 (2008).
\bibitem{5} M. H. Wood {\it et al.}, Phys. Rev. Lett. {\bf 105}, 112301 (2010).
\bibitem{6} M. Kotulla {\it et al.}, Phys. Rev. Lett. {\bf 100}, 192302 (2008).
\bibitem{7} H. Alvensleben {\it et al.}, Phys. Rev. Lett. {\bf 24}, 786 (1970).
\bibitem{8} M. Nanova {\it et al.}, arXiv: nucl-ex/1005.5694.
\bibitem{9} M. Nanova {\it et al.}, arXiv: nucl-ex/1008.4520.
\bibitem{10} M. Naruki {\it et al.}, Phys. Rev. Lett. {\bf 96}, 092301 (2006).
\bibitem{11} R. Muto {\it et al.}, Phys. Rev. Lett. {\bf 98}, 042501 (2007).
\bibitem{12} A. Polyanskiy {\it et al.}, Phys. Lett. B {\bf 695}, 74 (2011).
\bibitem{13} G. Agakishiev {\it et al.}, Phys. Rev. Lett. {\bf 75}, 1272 (1995).
\bibitem{14} M. Masera, Nucl. Phys. A {\bf 590}, 93c (1995).
\bibitem{15} G. Agakishiev {\it et al.}, Phys. Rev. C {\bf 84}, 014902 (2011).
\bibitem{16} D. Adamova {\it et al.}, Phys. Rev. Lett. {\bf 91}, 042301 (2003).
\bibitem{17} R. Arnaldi {\it et al.}, Phys. Rev. Lett. {\bf 96}, 162302 (2006).
\bibitem{18} S. Damjanovic {\it et al.}, Eur. Phys. J. C {\bf 49}, 235 (2007).
\bibitem{19} R. Arnaldi {\it et al.}, Eur. Phys. J. C {\bf 61}, 711 (2009).
\bibitem{20}R. S. Hayano and T. Hatsuda, Rev. Mod. Phys. {\bf 82}, 2949 (2010);\\
            S. Leupold, V. Metag, U. Mosel, Int. J. Mod. Phys. E{\bf 19}, 147 (2010).
\bibitem{21} M. Nanova {\it on behalf of the CBELSA/TAPS Collaboration},\\
              arXiv: nucl-ex/1109.4029; nucl-ex/1111.6006.
\bibitem{22} M. M. Kaskulov and E. Oset, AIP Conf. Proc. {\bf 842}: 483--485 (2006) [arXiv: nucl-th/0512108];\\
             M. M. Kaskulov and E. Oset, Phys. Rev. C {\bf 73}, 045213 (2006).
\bibitem{23} L. Tol$\acute{\rm o}$s {\it et al.}, Phys. Rev. C {\bf 82}, 045210 (2010).
\bibitem{24} M. Kaskulov, L. Roca, E. Oset, Eur. Phys. J. A {\bf 28}, 139 (2006).
\bibitem{25} E. Ya. Paryev, J. Phys. G. {\bf 37}, 105101 (2010).
\bibitem{26} C. Amsler {\it et al.} [Particle Data Group], Phys. Lett. B {\bf 667}, 1 (2008).
\bibitem{27} F. W. Wieland {\it et al.}, arXiv: nucl-ex/1011.0822.
\bibitem{28} N. Muramatsu {\it et al.}, Phys. Rev. Lett. {\bf 103}, 012001 (2009).
\bibitem{29} D. P. Barber {\it et al.}, Z. Phys. C {\bf 7}, 17 (1980).
\bibitem{30} A. M. Boyarski {\it et al.}, Phys. Lett. B {\bf 34}, 547 (1971).
\bibitem{31} Z. W. Zhao {\it et al.}, CP1257, Hadron 2009, Proc. of the {\it XIII} Int. Conf.\\
             Edited by V. Crede, P. Eugenio, and A. Ostrovidov (2010);\\
             AIP Conf. Proc. {\bf 1257}: 562--565 (2010).
\bibitem{32} H. Kohri {\it et al.}, Phys. Rev. Lett. {\bf 104}, 172001 (2010).
\bibitem{33} S. I. Nam, A. Hosaka, H. C. Kim, Phys. Rev. D {\bf 71}, 114012 (2005).
\bibitem{34} A. I. Titov {\it et al.}, Phys. Rev. C {\bf 72}, 035206 (2005).
\bibitem{35} A. Sibirtsev {\it et al.}, Eur. Phys. J. A {\bf 31}, 221 (2007).
\bibitem{36} S. I. Nam {\it et al.}, Phys. Rev. D {\bf 75}, 014027 (2007).
\bibitem{37} S. I. Nam and C. W. Kao, Phys. Rev. C {\bf 81}, 055206 (2010).
\bibitem{38} J. J. Xie and J. Nieves, Phys. Rev. C {\bf 82}, 045205 (2010).
\bibitem{39} CNS Data Analysis Center, The George Washington University, http://gwdac.phys.gwu.edu.
\bibitem{40} A. S. Iljinov {\it et al.}, Nucl. Phys. A {\bf 616}, 575 (1997).
\bibitem{41} O. Benhar {\it et al.}, Phys. Rev. D {\bf 72}, 053005 (2005);\\
             O. Benhar and G. Veneziano, arXiv: nucl-th/1103.0987.
\bibitem{42} A. Sibirtsev and M. B$\ddot{\rm u}$scher, Z. Phys. A {\bf 347}, 191 (1994).
\bibitem{43} V. K. Magas {\it et al.}, arXiv: nucl-th/0911.3614.
\bibitem{44} S. V. Efremov and  E. Ya. Paryev, Eur. Phys. J. A {\bf 1}, 99 (1998).
\bibitem{45} A. Zabrodin {\it et al.}, Phys. Rev. C {\bf 55}, R1617 (1997).
\bibitem{46} A. Braghieri {\it et al.}, Phys. Lett. B {\bf 363}, 46 (1995).
\bibitem{47} ABBHHM Collaboration, Phys. Rev.  {\bf 175}, 1669 (1968).
\end{thebibliography}
\end{document}